\documentclass{sig-alternate}

\usepackage{balance}
\usepackage{booktabs} 
\usepackage{cite}
\usepackage{color}
\usepackage{stackrel}
\usepackage{siunitx}
\usepackage{xspace}   
\usepackage{enumitem} 
\usepackage{url}

\newcommand{\Paragraph}[1]{\smallskip\noindent{\bf #1.}}

\newcommand{\eg}{e.\@\,g.\@\xspace}

\usepackage{listings}

\def\sharedaffiliation{%
\end{tabular}
\begin{tabular}{c}}

\makeatletter
\def\@copyrightspace{\relax}
\makeatother

\begin{document}

\title{MiniCPS: A toolkit for security research on CPS Networks}
%

\numberofauthors{2}
\author{
\alignauthor Daniele Antonioli\\
\email{daniele\_antonioli@sutd.edu.sg}
\alignauthor Nils Ole Tippenhauer\\
\email{nils\_tippenhauer@sutd.edu.sg}\\
\sharedaffiliation
\affaddr{Information Systems Technology and Design Pillar,}\\
\affaddr{Singapore University of Technology and Design}\\
\affaddr{8 Somapah Road}\\
\affaddr{487372 Singapore}\\
}

\maketitle
\begin{abstract}
  In recent years, tremendous effort has been spent to modernizing
  communication infrastructure in Cyber-Physical Systems (CPS) such as
  Industrial Control Systems (ICS) and related Supervisory Control and
  Data Acquisition (SCADA) systems.  While a great amount of research
  has been conducted on network security of office and home networks,
  recently the security of CPS and related systems has gained a lot of
  attention. Unfortunately, real-world CPS are often not open to
  security researchers, and as a result very few reference systems and
  topologies are available.
  
  In this work, we present \emph{MiniCPS}, a CPS simulation toolbox
  intended to alleviate this problem. The goal of MiniCPS is to create
  an extensible, reproducible research environment targeted to
  communications and physical-layer interactions in CPS. MiniCPS
  builds on Mininet to provide lightweight real-time network emulation,
  and extends Mininet with tools to simulate typical CPS components
  such as programmable logic controllers, which use industrial protocols
  (Ethernet/IP, Modbus/TCP). In addition, MiniCPS defines a simple API
  to enable physical-layer interaction simulation. In this work, we
  demonstrate applications of MiniCPS in two example scenarios, and
  show how MiniCPS can be used to develop attacks and defenses that
  are directly applicable to real systems.

\end{abstract}



\section{Introduction}
Industrial Control Systems (ICS) and Supervisory Control and Data
Acquisition (SCADA) systems traditionally relies on communication
technology such as RS-232 and RS-485, and field buses such as
Profibus. Due to the long lifetime of industrial components in such
settings, transitions to technology such as Ethernet, TCP/IP, and
related protocols are only implemented now. The adoption to the
standard internet protocol suite is expected to enhance
interoperability of the equipment, and reduce overall communication
costs.


The growing connectivity is also expected to introduce novel security
threats, in particular when systems are communicating over public
networks such as the internet. While a great amount of research has
been conducted on network security of office and home networks,
recently the security of CPS and related systems has gained a lot of
attention~\cite{chabukswar2010simulation,LinYangXuZhao,wangYummingXiaofeiYiuHuiChow,zhuJosephSastry,zonouzRogersBerthierBobbaSandersOverbye}. Unfortunately,
real-world CPS are often not open to security researchers, and as a
result no reference systems are available. In addition, physical layer
interactions between components need to be considered besides network
communications. We believe that this will require novel simulation
environments, that are specifically adapted to cater for the
requirements of CPS and ICS.

In this work, we present \emph{MiniCPS}, a CPS simulation toolbox
intended to alleviate this problem. The goal of MiniCPS is to create
an extensible, reproducible research environment targeted towards
CPS. MiniCPS will allow researchers to emulate the network of an
industrial control system, together with simulations of components
such as PLCs. In addition, MiniCPS supports a basic API to capture
physical layer interactions between components. Based on MiniCPS, it
is possible replicate emulate ICS in real-time, for example to develop
novel intrusion prevention systems, or own software to interact with
industrial protocols. While not all CPS systems are using
Ethernet-based communication so far, we see a general trend towards
wide adoption of Ethernet, even in applications such as airplanes,
vehicles, and embedded systems.

MiniCPS can also be used to share different system setup easily, and
can be extended by standard Linux tools or projects. Due to our use of
Mininet for the network emulation part, MiniCPS is especially well
suited to perform research on Software-Defined Networking in the
context of Industrial Control Systems.

We summarize our contributions as following:
  \begin{itemize}[noitemsep,nolistsep]
  \item We identify the issue of missing network simulation
    environments for applications such as cyber-physical systems. In
    particular, such simulation environment should support physical
    interactions, detailed communication links, and specific
    industrial protocols that are used.
  \item We present MiniCPS, a framework built on top of Mininet, to
    provide such a simulation environment.
  \item We present an example application cases in which we use
    MiniCPS to develop and refine a specific attack, which we later
    validated in a real testbed.
  \item We propose the use of Software-Defined Networking for CPS
    networks to enable efficient detection and prevention of the
    attack presented earlier. We design an implement a matching
    controller in MiniCPS.
  \end{itemize}

  The structure of this work is as follows: In
  Section~\ref{sec:background}, we introduce Mininet and CPS networks
  in general. We propose our MiniCPS framework in
  Section~\ref{sec:minicps}, and provide an application example in
  Section~\ref{sec:examples}. In Section~\ref{sec:sdn}, we show how
  MiniCPS can be used to develop a CPS network specific SDN
  controller.  Related work is summarized in
  Section~\ref{sec:related}. We conclude the paper in
  Section~\ref{sec:conclusions}.

\section{CPS Networks and Mininet}
\label{sec:background}
In this section, we will introduce some of the more salient properties
of industrial control system (ICS) networks that we have found so far. In
addition, we will briefly introduce Mininet, the network simulation
tool we use as part of MiniCPS.

\subsection{ICS networks}

In the context of this work, we consider industrial control systems
that are used to supervise and control system like public
infrastructure (water, power), manufacturing lines, or public
transportation systems. In particular, we assume the system consists
of programmable logic controllers, sensors, actuators, and supervisory
components such as human-machine interfaces and servers. We focus on
single-site systems with local connections, long distance connections
would in addition require components such as remote terminal units
(see below). All these components are connected through a common
network topology.

\Paragraph{Programmable logic controllers} (PLCs) are directly
controlling parts of the system by aggregating sensor readings, and
following their control logic to produce commands for connected
actuators.

\Paragraph{Sensors and actuators} are directly connected to the
network (or indirectly via remote IOs or PLCs).

\Paragraph{Network Devices} ICS often use \emph{gateway} devices to
translate between different industrial protocols (e.g. Modbus/TCP and
Modbus/RTU) or communication media. In the case where these gateways
connect to a WAN, they are usually called \emph{remote terminal units}
(RTUs).

\Paragraph{Network Topology} Traditionally, industrial control systems
have seen a wide deployment of direct links between components, based
on communication standards like RS-232. In addition, bus systems such
as RS-485 and Profibus have been used. In particular, focus on
reliability led to a wide deployment of topologies such as rings,
which could tolerate failure of a single component without loss of
communications, with very low reaction time (typically in the order of
milliseconds).

In recent years, industrial networks are transitioning to mainstream
consumer networking technology (i.e. Ethernet, IP, TCP). Nevertheless,
the need for reliability and interoperability with existing equipment
leads to use of additions that are uncommon in typical home and office
networks, such as Ethernet rings, use of IP-layer multicasting, and
custom protocols such as Ethernet/IP (ENIP). ENIP is an
application-layer protocol that transports \emph{Common Industrial
  Protocol} (CIP) messages that can be used to query sensor readings
from components, set configuration options, or even download new logic
on a PLC. In that model, sensor readings or control values are
represented by \emph{tags} (which can be roughly related to public
variables in programming). CIP uses a request-response model where a
client sends a request to a server (for example to read a \emph{tag}
containing a value read from a hardware component) and where the
server then sends back a reply (\eg with the requested value or an
error code). Such requests can operate on \emph{tags} and also on the
metadata associated with the tag, like access control and data type,
which are stored in \emph{attributes}. ENIP handles the \emph{session}
aspect of communications, for example with connected sessions (with
handshake and tear-down messages) and unconnected sessions (without
any handshake but with more contextual data in every CIP packet).

\Paragraph{Topology layers} Networks for industrial control systems
are often grouped in several layers (more detail on such networks
in~\cite{moyne07emergence}). On the lowest layer (which we call layer 0 or L0), sensors and
actuators are connected to controllers such as PLC. The sensors and
actuators are either capable of connecting to a network directly
(e.g., using ENIP), or they use basic analog or digital signaling,
which has to be converted to Ethernet-based communications by
\emph{remote input/output} (RIO) devices. Only if actuators and
sensors are physically very close to the PLC, the IO modules will be
installed as part of the PLC.

\begin{figure}[tb]
\centering
\includegraphics[width=\linewidth]{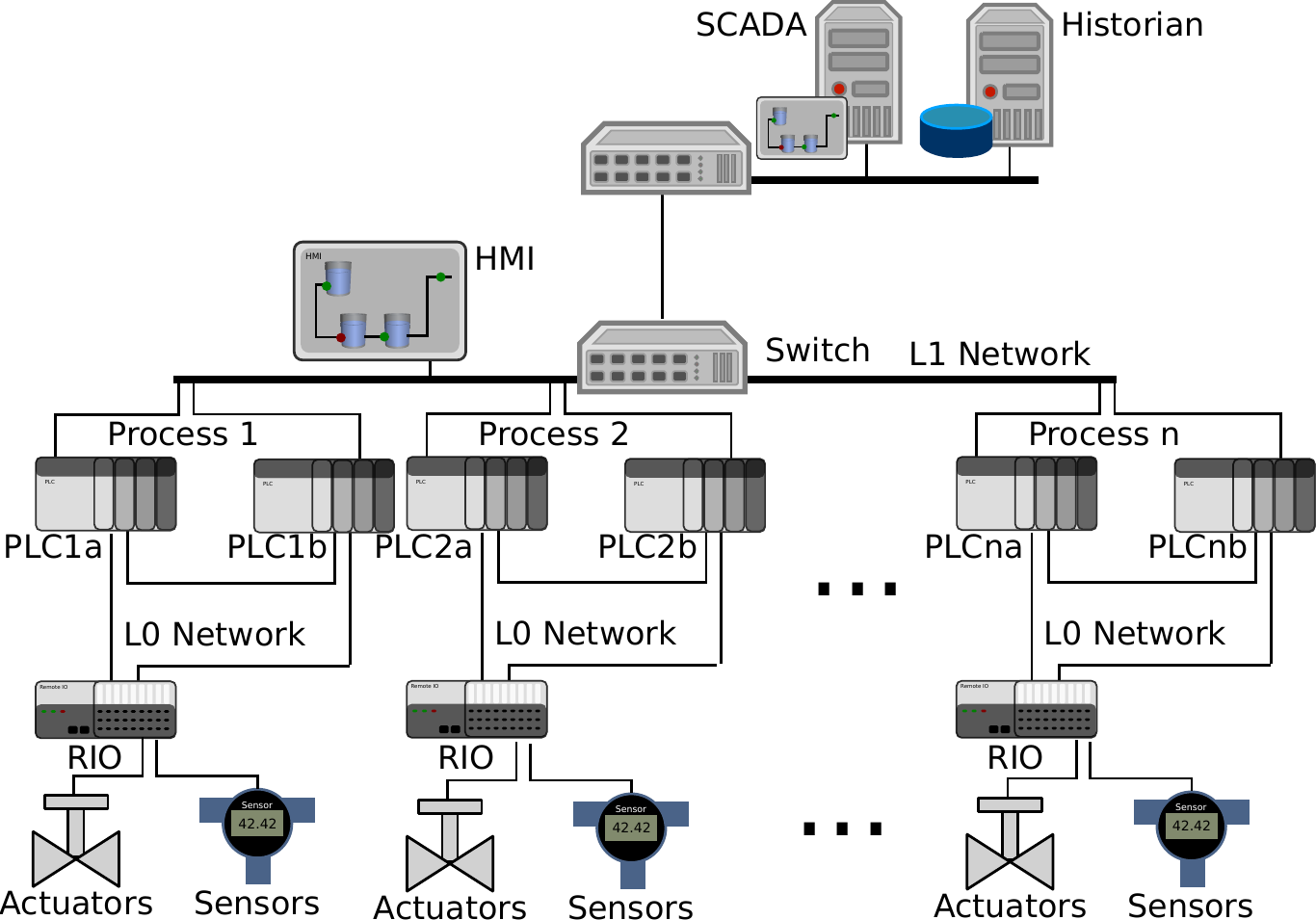}
\caption{Example local network topology of a plant control network.}
 \label{fig:generic}
\end{figure}

The next higher layer (layer 1/ L1) will connect the different controllers
(PLCs) with each other, together with local control such as
Human-Machine-Interfaces (HMI), local engineering workstations, and
Data historians. For simplicity, all these devices are often kept in
the same IP-layer subnetwork, although more complex topologies are
possible. We also note that industrial Ethernet switches are often
focused on electrical reliability, instead of IP-layer functionality
(e.g. the Stratix 5900 switch). We provide the network topology of a
generic ICS network as an example in Figure~\ref{fig:generic}.

\subsection{Mininet}
Mininet~\cite{lantz10Mininet} is a network simulator that
allows to emulate a collection of end-hosts, switches, routers, middle boxes,
and links with high level of fidelity.
It enables rapid testing and prototyping of large network setups on
constrained resources, such as a laptop. Furthermore, it was build around
Software-Defined Networking paradigm, facilitating SDN research and
development~\cite{oliveira14sdn}.

Mininet exploits lightweight system virtualization using Linux
\emph{containers}. A container can group a subset of processes and give
them independent view of system resources.
This approach presents various advantages over a full system
virtualization: Mininet runs on a single
kernel, its computational overhead is lower and the emulator can easily
tolerate scalability issues (e.g. one thousand containers instead of one
thousand dedicated virtual machines).

Each virtual host is a collection of processes isolated into a
container.  A \emph{virtual network namespace} is attached to each
container and it provides a dedicated virtual interface and private
network data. Link are emulated using virtual Ethernet (\texttt{veth})
and they can be shaped through Linux Traffic Control
(\texttt{tc}). Each virtual host utilizes its virtual interface to
send packets to a software switch.

Mininet can be used in multiple scenarios and can be easily adapted
over time to track the evolution of CPS networks. It provides a
realistic simulation environment to the user, and one can work with
the same addresses, protocol stacks and network tools of a physical
network, it is even possible to reuse helper scripts and configuration
files from the simulated environment directly in the physical network.

Mininet ships with a set of prepared topologies, in addition the user
can easily extend this collection through the provided Python
APIs. Dynamic interaction within any chosen topology can be achieved
through a convenient command line interface. Mininet is free,
open-source, well documented and actively maintained by a strong and
competent community. Furthermore, Mininet gives the opportunity to the
user to develop OpenFlow network architectures with transparent
integration of experimental code into production code.

\section{MiniCPS}
\label{sec:minicps}

In this section, we will introduce \emph{MiniCPS}. MiniCPS provides a
set of Python tools to enable real-time emulation of network traffic
in CPS such as ICS. This emulated system will allow (a) researchers to
build, investigate, and exchange ICS networks, (b) network engineers
to experiment with planned topologies and setups, and (c) security
experts to test exploits and countermeasures in realistic virtualized
environments.

In MiniCPS, components such as PLCs are emulated by python scripts
that manage the decoding of industrial protocols and physical layer sensors
and actuators signals. All networked system components (including
switches) are emulated using Mininet, discussed in detail in
Section \ref{sec:background}. Physical layer interactions
are currently modeled by a simple API (based on shared read/write to
files).

\subsection{Goals of MiniCPS}
In addition to the general application goal as outlined above, our
design of the MiniCPS toolkit is based on the following goals.
\begin{itemize}[noitemsep,nolistsep]
\item Cost-effectiveness (in particular, compared to real testbed)
\item Compatibility (you can deploy results on hardware)
\item Realistic simulation of industrial traffic (e.g., ENIP)
\item Open-source licenses (research friendly)
\item Future readiness (support application of SDN to CPS)
\item Usability: the tool should be easy configure (API hides low-level details)
\item Reproducibility: the tool should enable easy sharing of results
  between users
\end{itemize}

While most of these goals should be quite intuitive, we will comment on selected ones in the following.

\Paragraph{Reproducibility}
In~\cite{handigol12containers}, the authors proposed to use tools such
as Mininet to disseminate reproducible research results. In
particular, researchers can make the scripts to generate their network
setups public, which allows other researchers to reproduce the exact
same environments for their experiments. We strongly believe that such
dissemination of results would also be helpful in the context of
security research, in particular when systems which are less
mainstream are considered. While it is relatively easy to replicate
office network settings as related software is well-known, specialized
application setups such as ICS would be valuable to share.

\Paragraph{Compatibility}
We aim to provide a platform that allows direct application of
standard networking tools, as well as applications designed for the
target CPS. In particular, we aim to not only provide an abstraction
of the network to perform simulations on (similar to network
simulators such as NS2~\cite{ns2intro}, Omnet~\cite{varga2001omnet}),
but we target a network emulation that is largely identical to a real
network, without the cost or overhead of running a real network or a
set of virtual machines. In particular, this would allow us to develop
components that are directly using industrial protocols to
communicate. In addition, detailed network emulation will allow us to
use novel concepts such as software defined networking in the context
of CPS networks (see Section~\ref{sec:sdn}). We note that to achieve
this compatibility, we will be constrained to real-time simulation
instead of being able to simulate with arbitrary speedup.

\Paragraph{What MiniCPS does Not aim for}
MiniCPS does not aim to be a performance simulator, or tool for
optimizations. In addition, we currently put very little emphasis on
GUI or visualization. We note that building on top of the physical
layer API, and by extending the component logic scripts in general, it
should be possible to easily create real-time charts of physical
process parameters or controller states. 

\begin{figure}[tb]
    \centering
    \includegraphics[width=0.55\linewidth]{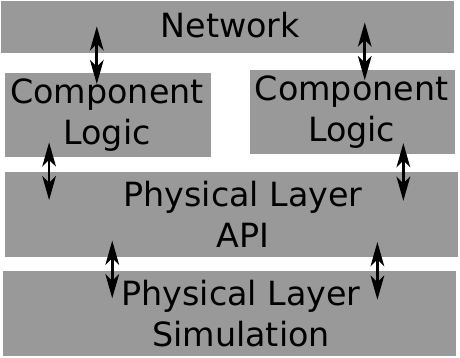}
    \caption{MiniCPS framework layers: CPS components are simulated as component logic, connected through the network emulation, and physical layer simulation.}
    \label{fig:blockscheme}
\end{figure}

\subsection{Design overview}
Components in MiniCPS interact on several layers (see
Figure~\ref{fig:blockscheme}). On the top layer, we have the network
through which messages are exchanged on top of ENIP, or other
protocols. Connected to this network are components, their logic is
implemented in simple scripts or more advanced software packages. If
the real-world counterpart of these components is interacting with the
physical layer, the simulated components will also have access to
specific physical layer properties through a second API, which
abstracts the physical layer. To simulate chemical or physical
processes, a selection of their properties are made available through
the API, and updated in real-time by simulation scripts. 

\subsection{Network Communication}
For the main network emulation layer of MiniCPS, we are using Mininet
(see Section~\ref{sec:background}). Mininet allows basic properties
such as delay, loss rates, and capacity of all links. In MiniCPS, we
use this functionality to allow individual links to be configured with
individual settings. As a result, we can emulate wide area network
connections and local are network connections with different
properties easily.

Based on Mininet, the network communication in MiniCPS uses the
default Linux networking stack based on Ethernet. All components have
virtualized network interfaces that are connected to each other. In
particular, this setup allows us to construct arbitrary topologies
such as simple star topologies of switches connected to devices,
intermediate routers and firewalls, and topologies such as Ethernet
rings. Protocols such as the spanning-tree-protocol or other routing
algorithms can be used to automatically avoid looping configurations,
and to establish routes. All standard protocols such as ICMP, HTTP,
NTP, etc. can be used right away.  On top of that, specific industrial
protocols can be used. In particular, we use the CPPPO Python library
to provide fundamental Ethernet/IP (ENIP) services~\cite{cpppo}. In
addition to ENIP, CPPPO also supports protocols such as Modbus/TCP. In
addition to CPPPO, we also use the pycomm library for ENIP
communications~\cite{pycomm}.



\subsection{Physical Layer Interactions}
Physical layer interactions between different components in the
systems are captured by our PHY-SIM API. This API is essentially a set
of resources (currently files), that provide data in real-time. These
resources can be read by components (i.e. a sensor reading some
physical property), or written to (typically, by a script that
emulates physical processes). The main purpose of the simple API is to
allow different tools to interact with it as easily as possible,
e.g. Matlab, python scripts, or dedicated physics
simulators. Representing the physical layer properties as file
resources makes this API independent of any particular library or
programming language. The files contain JSON data structures, which
are easy to parse and update. We also envision that it is possible to
connect these files to an actual physical process, i.e. to have the
process \emph{in the loop} (if suitable interfaces to the physical
system are provided). In the long term, the simple API could be
extended to a more generic API, for example a RESTful API using HTTP.

\subsection{Implementation}
MiniCPS is essentially a set of tools that extends Mininet, in
particular by adding simulation scripts for components such as PLCs,
HMIs, and historians, and by adding the physical layer API and
simulation part. As a result, the network emulation layer is built on
top of Mininet APIs. Our class hierarchy follows Object Oriented
design principles: every reusable, self-contained piece of code is
wrapped inside a class (such as a topology, a topology manager or an
SDN controller).

Our implementation contains three core modules: constants, topologies,
and devices. The \emph{constants} module collects data objects and
helper function common to all the codebase. The \emph{topologies}
module is a collection of ad-hoc CPS and ICS topologies with realistic
addresses and configurable link performance. The \emph{devices} module
contains a set of control logic applications developed using the pox
platform. Each core module is mirrored with a testing module
counterpart (even the constants).  Our class hierarchy design easily
allows Test Driven Development because each topology manager
potentially can select a network configuration, a controller, the
performance of the virtual links and even the CPU allocation for each
virtual host. In other words, a topology manager it is a
self-contained topology test. Indeed each test module is a collection
of \emph{test\_Something} classes with appropriate fixtures e.g. set
the Mininet log level at setup and clean Mininet containers at
tear-down.

We used the Python \emph{nosetests} module to automate test design, discovery,
execution, profiling and report. The \emph{logging} module enables
interactive code debugging/alerting and long time information storage.
Each core module and its testing counterpart append information to the
same log file, that rotates automatically through five time-sorted
backups.  SDN controllers log on separated files that are
(over)written at runtime. SDN code integration is obtained by means of
soft links using an initialization bash script.

We have implemented a first prototype version of MiniCPS, and are
currently in the process of testing and extending its
functionality. We plan to release the tool to the public in the near
future, using an open source license. All extensions are using the
Python programming language, and are documented using the Sphinx
package.

\section{Example Application: MitM traffic manipulations}
\label{sec:examples}
We mainly use MiniCPS to model the communications and control aspects
of a water treatment testbed at our institution. While the testbed is
intended for security research, we find it useful to have the MiniCPS
emulation environment to replicate the network settings outside the
lab. In addition to simulated interactions with PLCs and sensors, the
MiniCPS model also allows us to experiment with different network
topologies, and test SDN-related prototypes. In the following, we
highlight two such projects based on the MiniCPS model of our
testbed. The first application aims to provide on-the-fly manipulation
of ENIP/CIP traffic to change commands and sensor values as exchanged
between an HMI and a PLC. The second application (in
Section~\ref{sec:sdn}) concerns SDN controller-based detection and
mitigation of ARP spoofing attacks in the testbed.

\subsection{Basic Attack scenario}
ARP spoofing is a well-known attack in computer
networks~\cite{whalen2001introduction}. The attacker is connected to
the same Link Layer network segment as two victims, that are
exchanging messages. The attacker then sends specifically crafted
address resolution protocol (ARP) packets to both victims to cause
them to send their messages to the attacker, instead of each
other. The attacker then forwards the redirected messages to the
original recipient, which allows him to perform a stealthy
man-in-the-middle attack. We will show a possible countermeasure
against this attack in Section~\ref{sec:sdn}.

\begin{figure}[tb]
\centering
\includegraphics[width=\linewidth]{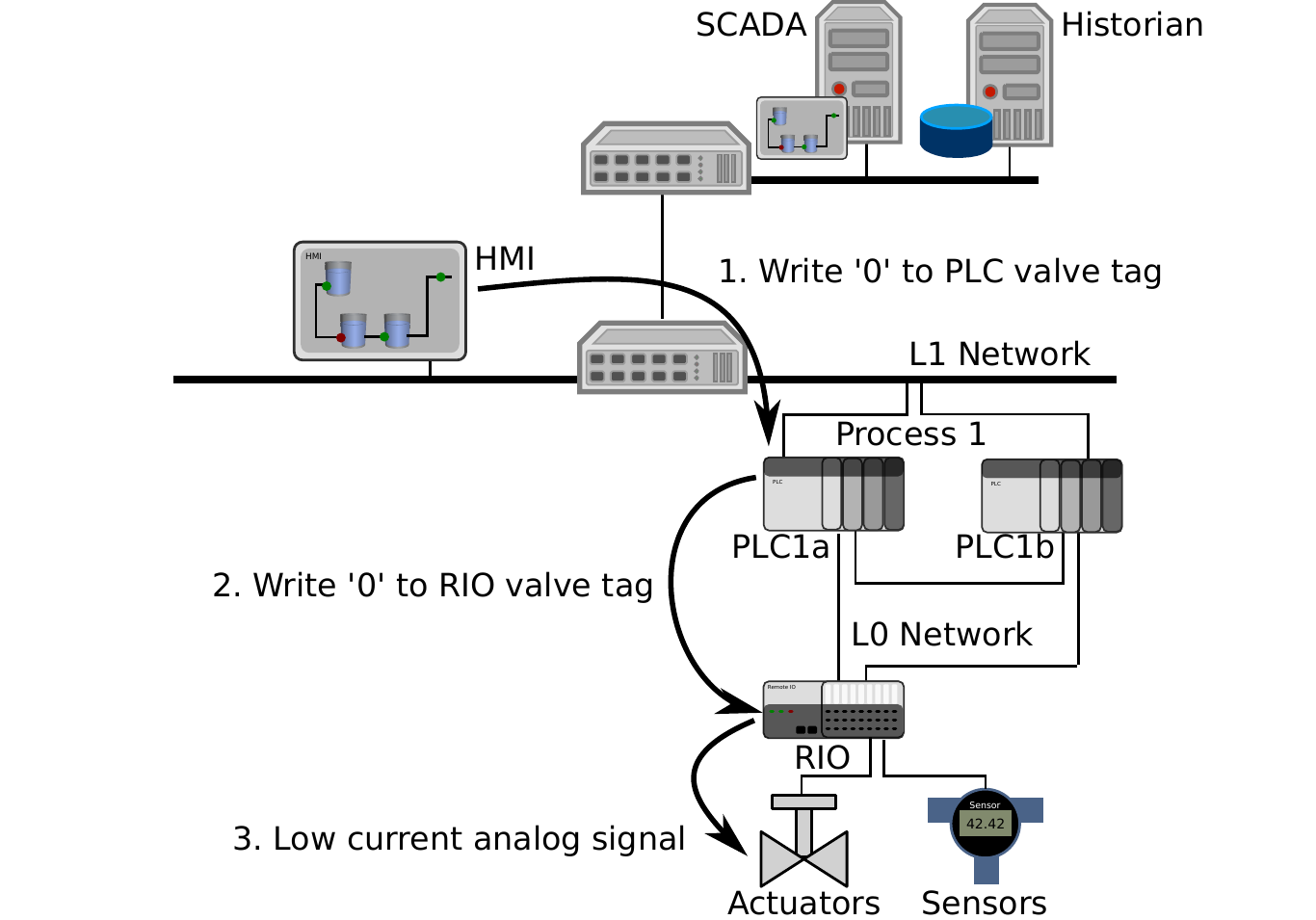}
\caption{Normal control message flow in the CPS. We omit the
  acknowledgment reply from the PLC in this visualization.}
 \label{fig:normalControl}
\end{figure}

Using ARP-spoofing, an attacker in the Layer 1 network of an ICS
system (see Figure~\ref{fig:normalControl}) can redirect all traffic
between two victim, e.g. PLC1 and the HMI. Let us assume the following
setting: the HMI is used to manually control the valve of a water feed
line towards a water storage tank. The control decision is done on the
HMI (e.g. operated by a human), based on the fill-level of the tank as
reported by a sensor in the tank. In this setting, the attacker now
aims to arbitrarily change the fill state of the tank, e.g. by filling
it over allowed maximal capacity, without being detected.

Based on that scenario, we modeled the network, HMI, PLC, and the
physical layer interaction between the valve and the tank in
MiniCPS. In particular, we modeled the valve as a simple Boolean
value, and the fill-state of the tank as a normal integer number.  The
valve value is periodically read by a process simulation script. If
the valve is open, the current fill-state of the tank is increased by
a fixed amount. Both the valve and fill-state are also used by the PLC
simulation script, which periodically reads the fill-state and
provides it as read-only CIP \emph{tag} to the emulated network. The
simulated PLC also provides a writable CIP tag for the valve
control. 

In practice we found that such settings are common. An attacker could
potentially overwrite the valve control tag (as there is no direct
access control in ENIP), but the HMI will continuously overwrite the
setting to its intended state (in our system, with 10Hz). As a result,
to continuously change the valve setting, the attacker has to send a
large amount of traffic to compete with the intended control by the
HMI, potentially interrupting normal operations. We developed an
alternative attack that does not increase the traffic load on either
HMI or PLC, and without interfering with other data exchanged between
PLC and HMI.

\subsection{Basic Attack}
\begin{figure}[tb]
\centering
\includegraphics[width=\linewidth]{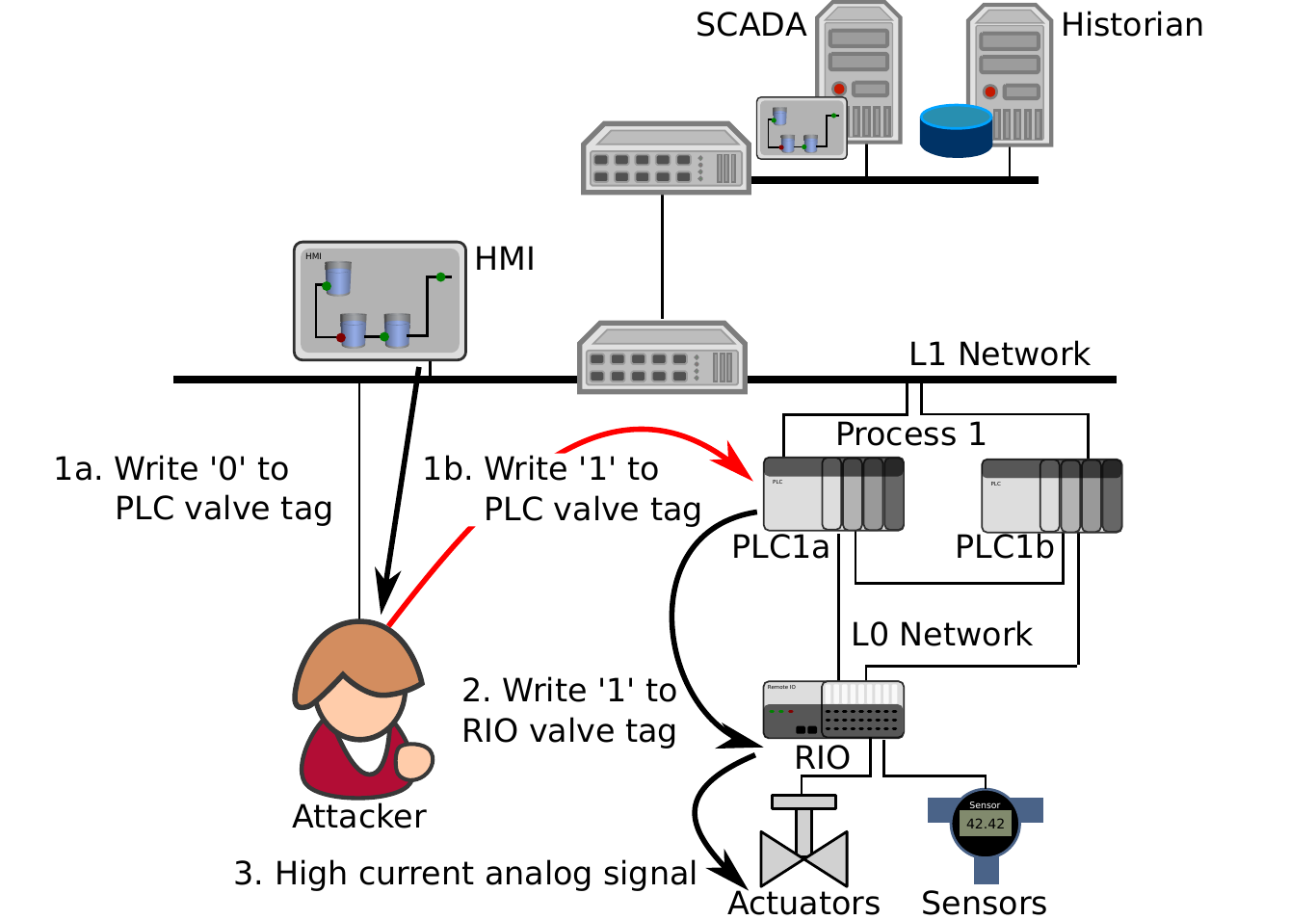}
\caption{Control message flow during the ARP spoofing attack.}
 \label{fig:arpControl}
\end{figure}

In a first simple attack (see Figure~\ref{fig:arpControl}), we used ettercap to install the attacker as
man-in-the-middle between the HMI and the PLC. We then wrote a set of
ettercap filter rules to change the value written by the HMI to the
valve tag at the PLC. As a result, each time the HMI sent a control
message to the PLC to keep the valve closed, the attacker could then
change this setting to ``open'', without fearing the HMI from
overwriting it again. We developed and deployed this attack in MiniCPS,
and were able to successfully change the valve tag to arbitrary values
as attacker.

\subsection{Simulating physical layer}
In our MiniCPS setup, we also simulated physical layer interactions as
outlined above. As result, the valve opened by the attacker led to an
increasing fill-state of the tank, which was in turn reported by the
PLC when queried by the HMI. In practice, this would allow the HMI to
at least trigger an alarm condition after the tank is exceeding the
maximal fill state. To prevent this detection, we extended our attack
by a seconds set of filter rules in the attacker. In addition to
rewriting the valve control values, the attacker now also rewrote the
value of the fill-state tag as reported from the PLC to the HMI. In
particular, the attacker could set this value to a constant, or apply
some noise to it if wanted. We successfully applied this attack in the
MiniCPS environment
. Afterwards, we
were able to apply the same attack to the real physical testbed, with
only minor modifications. The modifications were necessary as the
exact CIP messages exchanged between the HMI and PLC in the physical
testbed are not yet fully identical to the ones exchanged in our
MiniCPS environment.

\begin{figure}[tbh]
\centering
\includegraphics[width=0.8\linewidth]{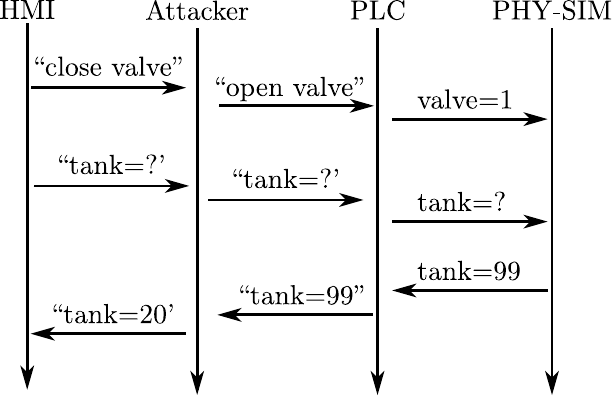}
\caption{Abstract messages in the extended attack: in addition to the
  modification of the control messages, the affected measurements from
  the PLC are also manipulated to hide the attack. In this setting, PHY-SIM could either be a real physical process, or our simulation layer.}
 \label{fig:attack2}
\end{figure}

\section{Example Application: SDN}
\label{sec:sdn}
There are a number of known countermeasures against the ARP spoofing
attack from the previous section (e.g., static ARP tables in the
hosts, traffic monitoring with an IDS). In the context of this
project, we were interested to see how a customized software-defined
network (SDN) controller could be used to detect and prevent the
attack outlined in the previous section. We now introduce SDN in
general, the POX controller project in particular, and then show how
we used MiniCPS to prototype a simple POX controller design
to prevent such ARP spoofing completely in our testbed.

In~\cite{dong15sdn}, the authors have presented a number of
motivations to use SDN in the context of smart power grid
communications. We compare our work with that work in more detail in
Section~\ref{sec:related}.
In a more general context, related work was published
recently in~\cite{zaalouk14orchsec,you14openflow}.


\subsection{Background on SDN/OpenFlow}
Software Defined Networking (SDN) is a novel
architectural way to think about building networks and OpenFlow is the
de-facto standard interface protocol between the SDN controlling logic
and the network devices (physical and virtual).  Both ideas were
proposed by \textsc{M. Casado} and they derives from
SANE~\cite{casado06sane}, a protection architecture for enterprise
networks.

The implementation defines a set of abstractions to provide separation
of concerns at the control plane, in a similar way as the layering
model that is used at the data plane. At the bottom of the stack there
are network devices that form the physical topology.  On top of that there
is a Network Operating System (NOS) able to talk to each device and to
serve a network view, in the form of an \emph{annotated graph}, to the
layer above. A virtualization layer is able to process this graph and
provide only relevant details to the level above through an API. At
the top of the stack there is the control logic that defines policy
over the network assessing the processed graph.  Communications
between the control logic and the physical devices is bi-directional:
network device messages will update the network graph and control
plane messages will update the network policy.  With this setting the
end-to-end principle, that again comes from the data plane management,
is reinforced also for the control plane. The (complex) management of
the network is shifted on the edges and central network devices merely
act as relays, becoming an homogeneous set of forwarding objects
referred as \emph{datapaths}.

In practice, in software defined networks, messages from the
switches (e.g. sent using OpenFlow) will be processed by a
\emph{controller}. For example, when a switch encounters a new flow
(e.g. a TCP connection with new target or new source), it will report
this flow to the controller via OpenFlow. The controller will then
analyze the flow, and informs the switch about appropriate actions to
take for the received messages (e.g. forwarding to a certain
port). Such controllers are realized by several
open source software projects.


\subsection{Leveraging SDN in CPS Networks}
\Paragraph{Why SDN for CPS Networks} The SDN paradigm presents some
interesting new possibilities when applied to CPS network design. The
control plane abstractions allow the designer to concentrate on the
network policy design. In addition, it is easier to develop, debug and
compare various control programs according to the requirements
(e.g. routing, isolation, traffic engineering). For further motivation
of SDN in the context of smart power grids, we refer
to~\cite{dong15sdn}. 

While in many applications, SDN is used to address highly dynamic
network conditions, traffic in industrial control systems is usually
quite predictable. In particular, topologies and the set of hosts
remain static (until the system is updated with new components). In
addition, we noticed that components exchange the essentially the same
traffic (with varying data payload of course). For example, tag values
could be queried every 100ms, and control commands could be sent every
second, resulting into regular traffic patterns. In the following, we
use the SDN paradigm to extract and enforce these traffic patterns,
which allows us to detect and prevent ARP spoofing attacks.

\Paragraph{SDN Controller Software} There are various interesting
projects regarding SDN and OpenFlow and it is relatively easy to find
a platform that implements the core modules, namely the NOS and the
virtualization abstractions. In our work we decided to use the
pox~\cite{pox} platform because it is targeted for the research
community, it offers out of the box libraries and components, and it
is object-oriented, event-driven with synchronous and asynchronous
handling capabilities. In addition, POX is completely written in
Python and it integrates well with our set of tools (scapy, cpppo,
Mininet, MiniCPS).


In a nutshell, events model communications
from the network to the controller (e.g. new datapath connection) and callback
functions model communications from the controller to the network
(e.g. add a new rule). In the next section we will introduce, as an illustrative
example, our ARP poisoning detection and handling scheme.

\begin{figure}[tb]
    \centering
   \includegraphics[width=\linewidth]{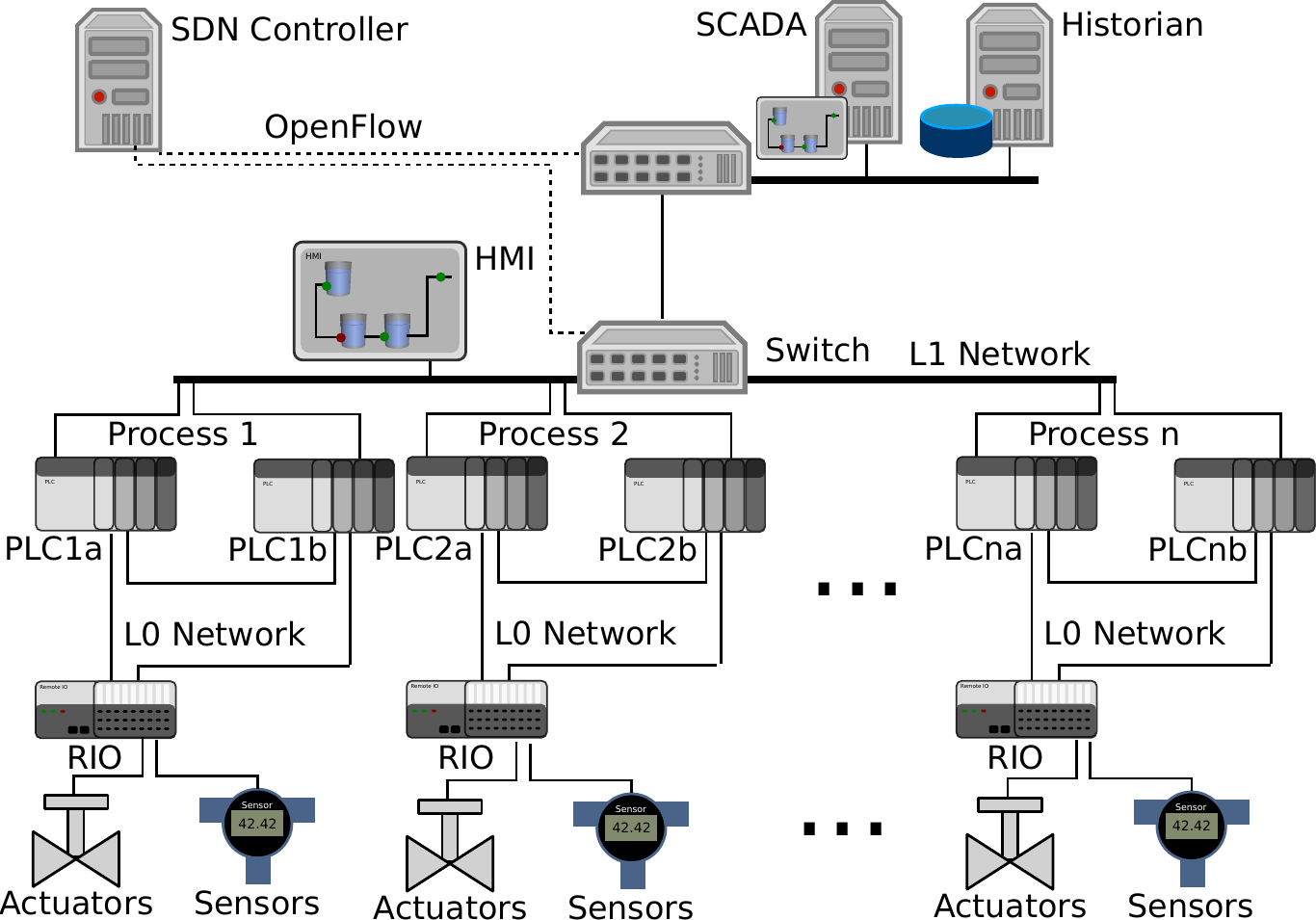}
   \caption{Extension of the generic ICS network with an OpenFlow
     switch and SDN controller.}
    \label{fig:sdnPox}
\end{figure}
\subsection{Preventing MitM attacks with a custom SDN controller}

We now present our SDN controller design, which aims to prevent the
ARP spoofing attacks as discussed in the previous Section. In
particular, our controller will analyze all ARP traffic, classify it
as malicious or benign, and then update the SDN switches with suitable
rules to prevent malicious attacks. Our threat model consists of an
attacker able to impersonate a CPS network device that aims to mount a
passive or active man-in-the-middle attack using ARP poisoning.

Our pox controller implements a fully centralized SDN
control plane with per-flow forwarding rules. Our control plane program uses
both a proactive approach to perform a static pre-mapping and a reactive approach
to adapt dynamically to the context. The detection and prevention code runs with higher
priority than the management code and it is able to block the event handling
chain.  

\Paragraph{Abstract overview} Every time a new switch is connected to the
network, our control logic will create a new reference to the network state
accessible by the switch. The network map comprises an \texttt{ip\_to\_mac}
and \texttt{mac\_to\_port} Python dictionaries. According to OpenFlow protocol,
when a switch doesn't know how to forward a packet it sends (a part of)
it to the controller. Our control logic process ARP reply and ARP request
messages verifying their consistency according to the map. 

Currently, suspicious ARP request are signaled and logged as warnings. Suspicious
ARP replies are actively managed: let's say that an attacker wants to
impersonate a PLC in the network, then our arp handling will detect the
spoofing by telling to the relevant switch to block all the traffic coming
from the attacker port and with the attacker MAC. 

Our mechanism detects both internal and external ARP spoofing attempt
and prevents both passive and active ARP MITM attacks. Under normal
ARP request/reply circumstances our controller dynamically update the
map of the network. In Listing~\ref{arp-code}, we present an
extract of our ARP request detection code.

\begin{figure}[tb]
\begin{lstlisting}[caption={ARP request spoofing detection.},label=arp-code,language=Python]
if sender_ip in self.ip_to_mac:
  if sender_mac != self.ip_to_mac[sender_ip]:
    # Internal attack
    if sender_mac in self.ip_to_mac.values():
      for key, value in self.ip_to_mac.items():
          if value == sender_mac:
              attacker_ip = key
              break
      self.log_internal()
      return True
    # External attack
    else:
      self.log_external()
      return True
return False
\end{lstlisting}
\end{figure}

Our implementation defines a set of ad-hoc handling functions that are called
before the standard pox event handlers. Our switch potentially can redirect
traffic to a dedicated Intrusion Detection System system for deep packet
inspection. We are using permanent flow rules to model our static CPS testbed
configuration. Our code can easily be integrated on any other pox-based SDN
controller \emph{without} changing the standard control logic.


In addition to this simple attack detection and prevention strategy,
we are currently developing more elaborated ARP detection and
mitigation techniques, in particular (i) an \emph{ARP cache restoring}
handler, and (ii) spoofing detection based on \emph{static mapping} of
MAC/IP pairs. The ARP restoring feature periodically or asynchronously
sends ARP replies to potentially every host in the network forcing it
to update its ARP cache with fresh and consistent data.

The second technique is the \emph{strong static premap} method, which
allows the controller to send to every new datapaths a set of
predefined flow rules to speedup initial traffic congestion and policy
establishment (e.g. who can talk to who).  Eventually, this mechanism
can be extended a dynamic policy checker component, that is able to
validate and restore the correct network state requesting and
processing general and aggregated flow statistics directly from the
datapaths.

Lastly, we would plan to extend our current centralized design
into a more robust distributed scheme by using multiple synchronized
controllers able to tolerate single point of failure in the control
plane domain. 

\nocite{heller09switch}

\section{Related work}
\label{sec:related}

Security aspects of CPS have been discussed
in~\cite{LinYangXuZhao,wangYummingXiaofeiYiuHuiChow,zhuJosephSastry,zonouzRogersBerthierBobbaSandersOverbye},
in particular in the context of smart power grid infrastructure and
control.

In~\cite{dong15sdn}, Dong \emph{et al} propose a testbed that is
similar to our MiniCPS platform in several ways. In particular, they
propose to use Mininet as network emulation platform, a power grid
simulation server, and a control center simulation server. The
envisioned testbed uses Mininet to simulate delays related to dynamic
network reconfigurations in the case of failures. In general, the
authors just discuss the use case of the smart power grid, with
component such as sensors and actuators connected to a central control
via RTUs.

We note that MiniCPS differs from the testbed in~\cite{dong15sdn} in
several ways. Most importantly, MiniCPS' focus is on sharing
reproducible CPS network topologies, in particular related to
industrial control systems. MiniCPS focuses on using a set of PLC
simulation tools, that directly interact with the network traffic, and
the physical layer API. The physical layer API abstraction is not
present in~\cite{dong15sdn}, as the authors propose the use of a
powerful power-grid simulation tool (PowerWorld). In MiniCPS, the
(generic) API would allow to combine different types of physical layer
simulations (e.g., combining water flow, mechanical levers,
temperature transfer). Finally, the industrial protocol differs (ENIP
vs. DNP3). From~\cite{dong15sdn}, it seems that the proposed testbed
was not yet fully implemented.

In~\cite{chabukswar2010simulation}, a framework with similar intent as
MiniCPS has been proposed. The framework uses OMnet++ as network
simulation tool, and also features simulation of physical layer
(e.g. a chemical plants). The authors
simulated denial of service attacks on the sensor data, and the
resulting control actions. As OMnet++ was used for network
simulations, network communication was simulated as abstract messages
that were routed through components, instead of simulating the full
TCP/IP+industrial protocol stack. As a result, attacks such as our
MitM ettercap manipulation could not be simulated in detail
(i.e. considering all fields of the CIP/ENIP messages). On the other
hand, simulations like~\cite{chabukswar2010simulation} allow
to use timescales other than real-time.

On the topic of SDN, SANE~\cite{casado06sane} represents one the first
practical SDN-based solution for secure network design. The proposed
implementation already included common SDN core concepts like
centralized control logic, high level network policy design and easy
network scalability.

SDN and OpenFlow projects involved from the beginning both academia
and leading IT industries, that eventually found the Open Networking
Foundation (ONF).  There are several other recommended papers about
SDN~\cite{feamster13sdn, thenewstack15sdn, onf12sdn} and
OpenFlow~\cite{mckeown08openflow, you14openflow}.


\section{Conclusion}
\label{sec:conclusions}
In this work, we proposed MiniCPS, which uses Mininet together with a
physical layer API and a set of matching component simulation tools to
build a versatile and lightweight simulation system for CPS
networks. While currently the physical layer simulation is very
simplistic, we believe that our general framework will (a) researchers
to build, investigate, and exchange ICS networks, (b) network
engineers to experiment with planned topologies and setups, and (c)
security experts to test exploits and countermeasures in realistic
virtualized environments.

MiniCPS builds on Mininet to provide lightweight real-time network
emulation, and extends Mininet with tools to simulate typical CPS
components such as programmable logic controllers, which use
industrial protocols (Ethernet/IP, Modbus/TCP). In addition, MiniCPS
defines a simple API to enable physical-layer interaction
simulation. We demonstrated applications of MiniCPS in two example
scenarios, and showed how MiniCPS can be used to develop attacks and
defenses that are directly applicable to real systems.

\section{Acknowledgments}
We thank Nicolas Iooss for his support and contributions related to Ethernet/IP support in MiniCPS and the demonstrated attacks, and Pierre Gaulon for his help on the physical layer simulation.


\end{document}